\documentclass{llncs}
\usepackage{graphicx}
\usepackage{algorithm}
\usepackage{algorithmic}
\usepackage{subfigure}
\usepackage{amsmath} 
\usepackage{color} 
\usepackage{comment} 

\begin{document}
\title{3DGraCT: A Grammar-based Compressed Representation of 3D Trajectories  
\thanks{\scriptsize{This work was funded in part by EU H2020 MSCA RISE BIRDS: 690941; MINECO-AEI/FEDER-UE: TIN2016-78011-C4-1-R; MINECO-CDTI/FEDER-UE CIEN IDI-20141259;  MINECO-CDTI/FEDER-UE CIEN IDI-20150616; MINECO-CDTI/FEDER-UE INNTERCONECTA ITC-20161074; Xunta de Galicia/FEDER-UE ED431C 2017/58 and ED431G/01. }}
}

\author{Nieves R. Brisaboa\inst{1} \and Adri\'an G\'omez-Brand\'on\inst{1} \and Miguel A. Mart\'inez-Prieto \inst{2} \and Jos\'e R. Param\'a\inst{1}
}
\institute{
 Universidade da Coru\~na, 
  CITIC, Spain\\
\email{\{brisaboa, adrian.gbrandon, jose.parama\}@udc.es}
\and
Universidad de Valladolid, Spain,
	\email{migumar2@infor.uva.es}
 }

\maketitle

\begin{abstract}
Much research has been published about trajectory management on the ground or at the sea, but compression or indexing of flight trajectories have usually been less explored. However, air traffic management is a challenge because airspace is becoming more and more congested, and large flight data collections must be preserved and exploited for varied purposes. This paper proposes 3DGraCT, a new method for representing these flight trajectories. It extends the GraCT compact data structure to cope with a third dimension (altitude), while retaining its space/time complexities. 3DGraCT improves space requirements of traditional spatio-temporal data structures by two orders of magnitude, being competitive for the considered types of queries, even leading the comparison for a particular one.
\end{abstract}

\section{Introduction}

Geopositioned data is ubiquitously and continuously generated to describe different types of trajectories; e.g. routes of  professional transportation vehicles or our daily running paths. Obviously, large and varied trajectory datasets are being consolidated, and they are exploited for different and innovative purposes. Disregarding their final application, managing trajectory datasets poses many challenges that have attracted much research efforts.  

A prominent domain that demands efficient trajectory management is Air Traffic Management (ATM). ATM systems analyze very large flight-related data\-sets to make decisions to improve air traffic performance, reducing costs, or making safer and environmentally friendly airspaces. Currently, ATM services are evolving to support and leverage ``next generation'' technologies like {\em Automatic Dependent Surveillance-Broadcast (ADS-B)}. ADS-B is a surveillance technology in which aircrafts determine flight parameters (latitude, longitude, altitude, etc) via navigation systems, and broadcast them to ground stations, that then deliver this data to ADS-B providers; e.g. the OpenSky Network \cite{Schafer:2014}, that is the provider of the ADS-B datasets used in our experiments.

ADS-B has been progressively adopted by many aircraft manufacturers, and more ground stations have been deployed around the world. It has increased ADS-B coverage, and also the size of ADS-B datasets, whose storage and querying has become more difficult. Storage issues were first addressed using columnar compression \cite{6887350,SO6C}. Although their numbers are moderately successful, the resulting representations can not be efficiently queried. More recently, a compressed index for ADS-B (called ADS-BI) has been proposed \cite{8283602}. It performs block partitioning and stores descriptive metadata about the block to enable some types of queries. Block contents are then encoded by columns using universal compression (e.g. gzip or p7zip), reporting competitive numbers. Although ADS-BI resolves some type of queries by time or 2D-position, it does not support altitude-based searches, which is highly desirable for ATM systems; for instance, when a controller looks for aircrafts flying at certain flight level in a given region. 

Therefore, our main objective is to propose a data structure that allows 3D trajectories to be effectively compressed, and searches to be performed by time and/or any of the three positional dimensions.  It is not a new problem \cite{Deng2011}, and some researches have been previously published about 2D {\em (latitude, longitude)}, and 3D (including {\em altitude}) trajectory management. Data structures like 3DR-tree \cite{Vazirgiannis1998}, HR-tree \cite{NascimentoS98}, the MVR-tree \cite{PapadiasT01}, or PIST \cite{Botea2008} have been successfully used for many years, but currently show scalability issues when they are used to manage larger trajectory datasets. The Douglas-Peucker algorithm \cite{Douglas:1973:ARN} has been used to make trajectories more compact; other examples are dead reckoning \cite{TrajcevskiCSWV06}, TrajStore \cite{TrajStore} and Trajic \cite{Trajic}.

Our approach, called 3DGraCT, proposes a new compact data structure that stores and indexes 3D trajectories in compressed space. 3DGraCT enhances GraCT \cite{Nav16} to manage altitude information, and also to enable query resolution by this dimension. Our experiments, using different-size ADS-B datasets, show that 3DGraCT improves space requirements of traditional spatiotemporal data structures by two orders of magnitude, and competes with them in query performance, leading the comparison for queries asking for large time intervals.

\vspace{-0.2cm}
\section{Background}
\begin{figure}[t]
	\centering
	\includegraphics[scale=0.15]{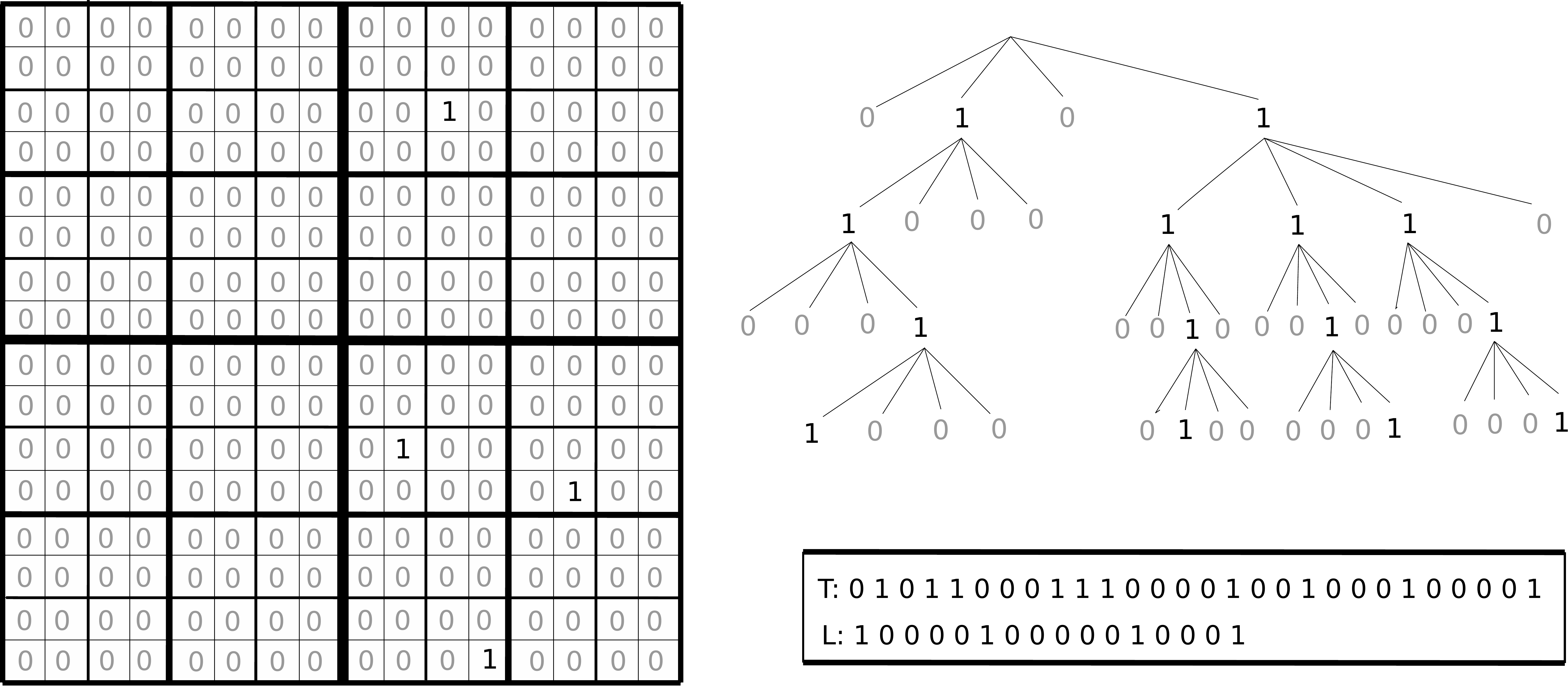} 
	\caption{\label{fig:example}Example of a binary matrix (left), the $k^2$-tree conceptual representation (top right), and the compact representation (bottom right), where $k=2$.}
\end{figure}

\noindent
{\bf $k^2$-tree}. The $k^2$-tree \cite{ktree} is conceptually an unbalanced $k^2$-ary tree constructed from a binary matrix by recursively subdividing the  matrix into $k^2$ submatrices of the same size, if $k=2$, it is a space/time efficient version of a region quadtree. First, the original matrix is divided into $k^2$ submatrices of size $n^2/k^2$, being $n \times n$ the size of the matrix.  Each of these submatrices  generates a child of the root node whose value is $1$, if there is at least one $1$ in the cells of that submatrix, and $0$ otherwise. The subdivision continues recursively for each child with value $1$ until a submatrix full of 0s is found or  the cells of the original matrix (i.e., submatrices of size $1\times 1$) ar reached. Figure~\ref{fig:example} shows an example of this subdivision (left) and the resulting conceptual $k^2$-ary tree (right up) for $k=2$. 

The $k^2$-tree is stored using two bitmaps $T$ and $L$ (see Figure \ref{fig:example}). \textit{T}  stores all the bits of the $k^2$-tree, except those in the last level,  following a level-wise traversal: first the $k^2$ binary values of the children of the root node, then the values of the second level, and so on. $L$ stores the last level of the tree. 


\medskip
\noindent
{\bf $k^3$-tree}. The $k^2$-tree can be generalized to deal with a
three-dimensional binary cube, instead of a two-dimensional binary matrix. It can be trivially done by extending the space partitioning, while maintaining the representation techniques used for $k^2$-trees.
Thus, each 1 in the binary cube of the $k^3$-tree \cite{k2ones} represents a tuple $\langle x,y,z \rangle$, where $(x,y)$ are the coordinates in the 2D space, and $z$ is the altitude. It is possible to obtain efficiently the value of a cell, a cube, or slices of the cube, by just performing $rank$ and $select$ operations \cite{Jacobson89} over 
$T$ and $L$.

\begin{figure}[t]
	\centering
	\includegraphics[scale=0.33]{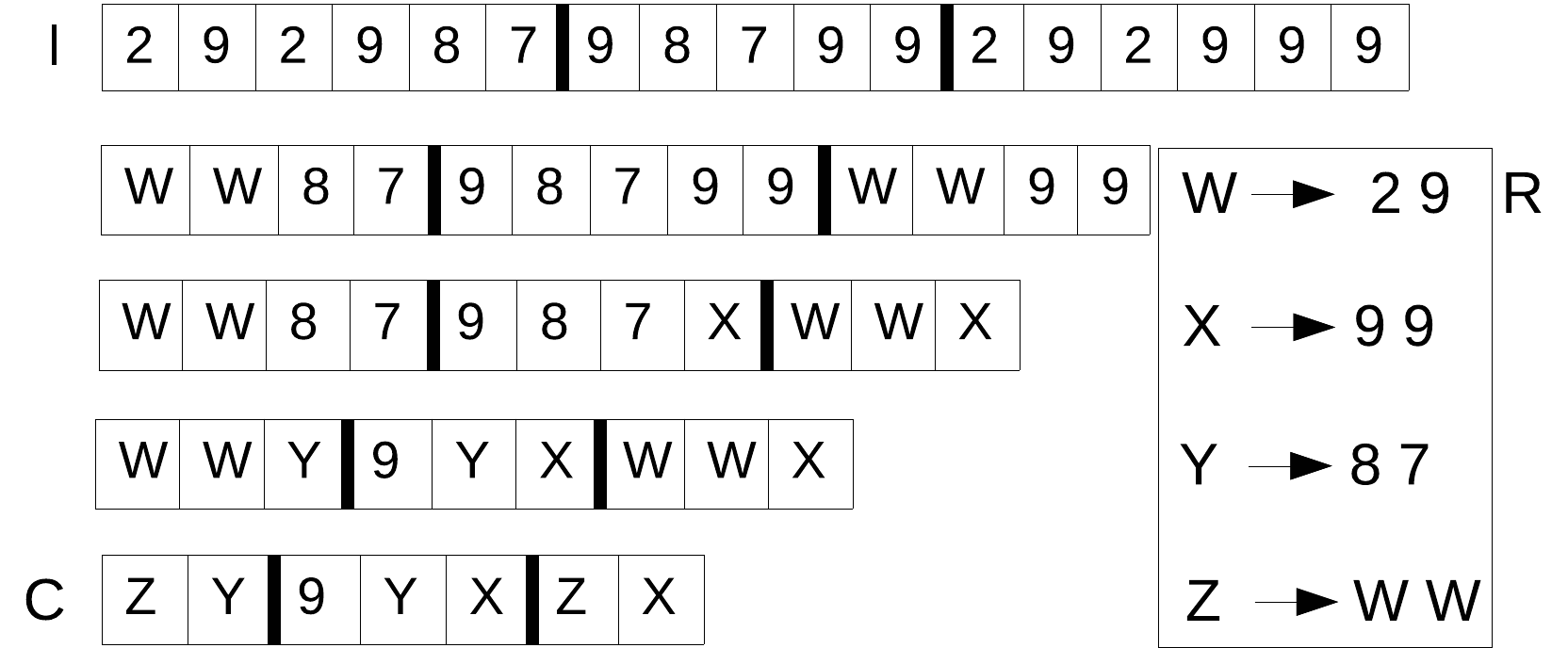}
	\caption{\label{re-pair}Example of Re-Pair compression.}
    \vspace{-0.4cm}
\end{figure}

\medskip
\noindent
{\bf Re-Pair.} Re-Pair~\cite{larsson2000off} compresses a sequence by recursively substituting pairs of symbols by a new one. Given a sequence of integers $I$ (called {\em terminals}) the compression process is as follows: (1) it obtains the most frequent pair of integers $ab$ in $I$; (2) it adds rule $W \rightarrow ab$ to dictionary $R$, where $W$ is a new symbol not present in $I$ (called a {\em non-terminal}); (3) every occurrence of $ab$ in $I$ is replaced by $W$, and (4) it repeats steps 1-3 until all pairs in $I$ appear only once (see Figure \ref{re-pair}). 
The resulting sequence after compressing $I$ is called $C$. 

\medskip
\noindent
{\bf GraCT.} GraCT \cite{RodriguezBrisaboa16} is a compact data structure to represent and query trajectories of moving objects in a free space of two dimensions. It requires that all objects declare their positions at regular time instants (e.g. each minute),  but interpolation is used  when an object does not inform its position in a given instant.
GraCT uses a raster model to represent the space; i.e. it is divided into cells (squares) of a fixed size, and it is assumed that objects fit in one of these cells. The size of the cells and the time elapsed between 
consecutive instants are parameters that can be adapted to particular cases. 

To store absolute positions of all objects, every $d$ time instants, GraCT uses a data structure based on the $k^2$-tree, which is called \textit{snapshot}. The distance, $d$, between snapshots is another parameter of GraCT. Between two consecutive snapshots, the trajectory of each moving object is represented as a \textit{log}, which is an array of relative movements with respect to the previous time instant.

\section{3DGraCT}
3DGraCT  proposes an extension of GraCT to three dimensions, so the space is divided into cells (small cubes) of fixed length, that form a bigger cube. 

\medskip
\noindent
{\bf Snapshots.} Each $d$ time instants, there is a snapshot ${\cal S}_{k}$, where $k$ is the time instant represented by the snapshot. These snapshots are organized as $k^3$-trees. A leaf of the $k^3$-tree set to 1 (i.e., a 1 in the bitmap $L$) means that one or more aircrafts\footnote{From now on, we will refer to them simply as objects or moving objects.} are placed in the corresponding cell, but the snapshot needs to determine which objects are located in that cell. Following the order of 1s in $L$, an array of object identifiers 
 (aircrafts) holds that information. This array is denoted as {\em perm}, since it is a permutation \cite{Knuth91a}. An additional bitmap, called \textit{Q}, is aligned with \textit{perm}. It marks with 0 that the aligned object identifier in \textit{perm} is the last object in the corresponding cell, and 1 means that more objects are located in that cell.

\begin{figure}[t]
\centering
\includegraphics[scale=0.22]{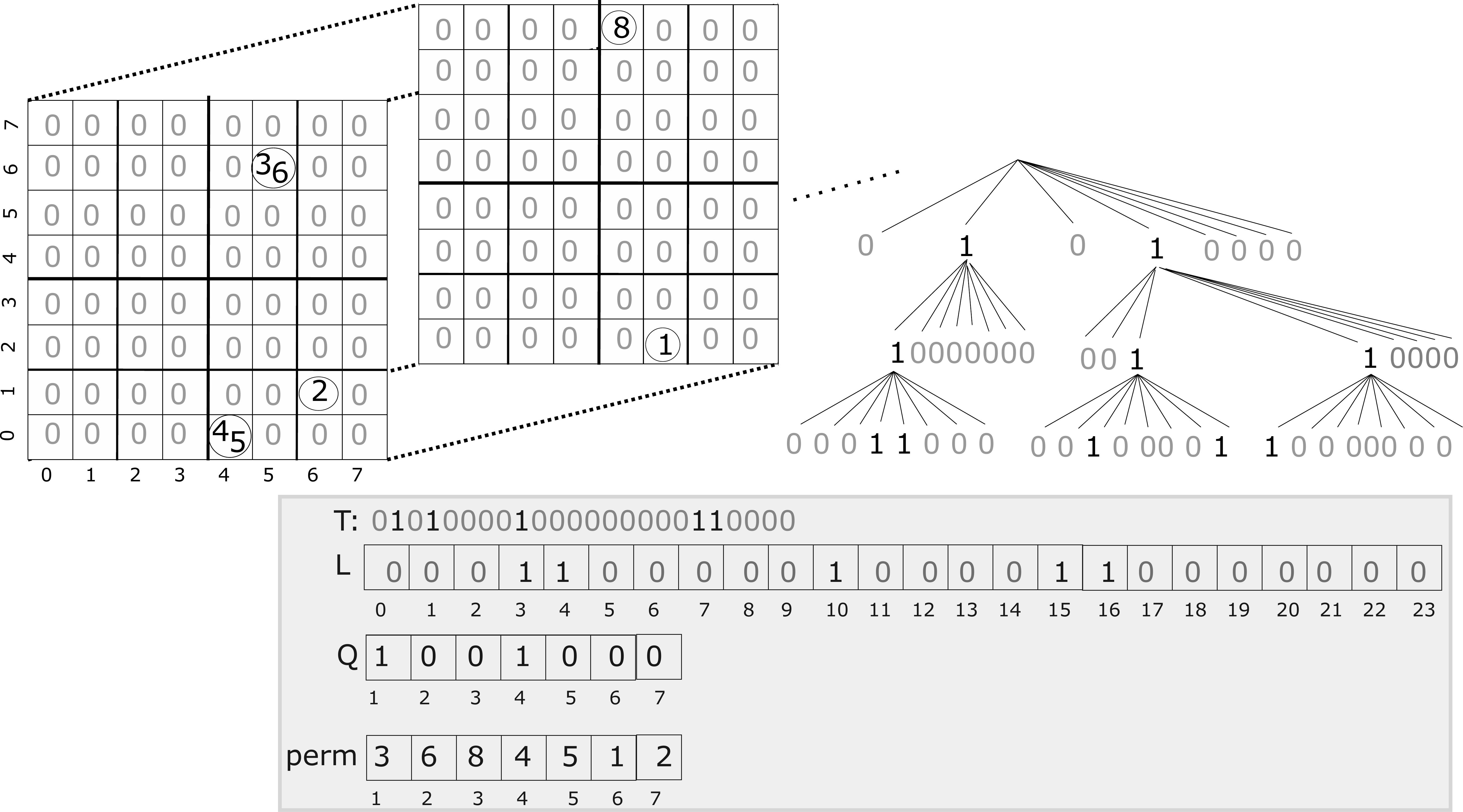}
\caption{The position of objects in the 3D space (top left), the conceptual $k^3$-tree (top right), and the snapshot (bottom).}\label{snapshot}
\end{figure}

Figure \ref{snapshot} shows an example of snapshot.\footnote{ Note that only shaded structures are used to encode the snapshot, the other ones are used for illustration purposes.} The two matrices models the first two slices of an $8 \times 8 \times 8$ cube representing the 3D space. Each slice contains the horizontal positions of all 
 aircrafts flying at a given altitude.  Each matrix shows object identifiers at certain positions, and the corresponding $k^3$-tree encodes this information by assuming that no objects are contained in the remaining slices. Each non-empty position in matrix corresponds to a bit set to 1 in $L$.
The object identifiers corresponding to the first 1 in $L$ (which is at position 3 of $L$) are stored starting at position 1 of $perm$.  $Q$ is then accessed to count the number of objects that are located in this cell: a sequential search is performed from $Q[1]$ until the first 0 (located at $Q[2]$). Thus,
there are two objects in the inspected cell.  The corresponding object identifiers are retrieved from {\em perm[1]=3} and {\em perm[2]=6}.
Now, in position 3 of {\em perm} starts the object identifiers corresponding to the second 1 in $L$, and so on.


 These structures allow 3DGraCT to address two types of queries:

\begin{itemize}
\item \textit{Find the objects in a box of the 3D space}. The $k^3$-tree is traversed from the root to the leaves to obtain
positions $n_1, n_2, \ldots n_m$, in $L$, that corresponds to positions  marked with 1 in the queried box. 
For each $n_i$, we count the number of 1s in the array of leaves $L$ until the position $n_i$; it obtains the number of  non-empty leaves 
up to the $n_i^{th}$ leaf, $x = rank_1(L, n_i)$. Then, the position of the $(x-1)$th 0 in $Q$ is obtained, which indicates the last bit of the previous leaf (with objects), and by adding 1, we get the first position in $perm$ with the objects of the  leaf corresponding to $n_i$, $p = select_0(Q, x-1)+1$. From $p$,
object identifiers aligned with 1s in $Q$ are retrieved, until a 0 is reached (it marks the last object identifier located in a leaf).

\item \textit{Find the position in the 3D space of a given object}.  The desired object identifier is first searched in \textit{perm}. Our permutation is enhanced with \textit{shortcuts} to avoid sequential searches. Assuming the object identifier is located at position $k$, the following step looks for its corresponding position in $L$.
We calculate the number of leaves before the object at {\em perm[k]}: 
$y = rank_0(Q, k-1)$. Then we find in $L$ the position of the
$(y+1)^{th}$ 1, that is, $select_1(L, y+1)$. This value is used to traverse the $k^3$-tree upwards in order to obtain the cell position in the 3D space, and thus the horizontal position and altitude of the object.
\end{itemize}

\begin{figure}[t]
\centering
\includegraphics[scale=0.19]{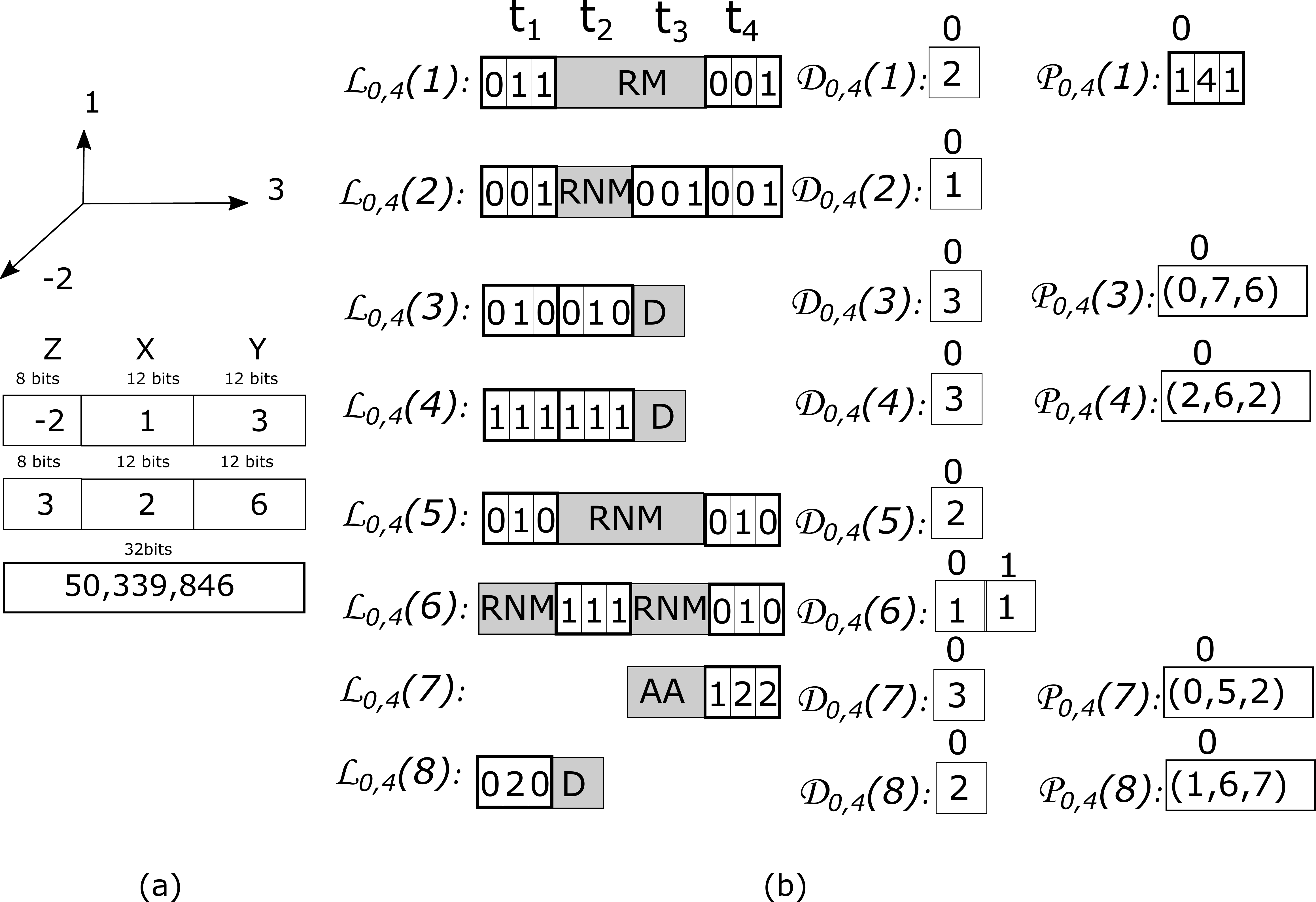}
\vspace*{-5mm}
\caption{The encoding of relative movements (left) and logs of objects (right).}\label{log}
\end{figure}

\medskip
{\bf Log of relative movements.}
The use of a snapshot for encoding each time instant would consume too much space, instead, between snapshots, 3DGraCT stores for each aircraft the  relative movements with respect to the last known position. A relative movement consists of 3 values, $\langle x, y, z \rangle$, which are the number of cells of difference between the new position and the last known position, in each dimension.
Probably,  $\langle x, y, z \rangle$ will be numbers with a small magnitude, as the differences between consecutive time instants cannot be very big. Instead of using 32 bits for each value,  we fit the three values into a 32-bit integer using 12 bits for the $x$ and $y$ values and 8 bits for the $z$ component. In Figure \ref{log}(a), we can see a relative movement of 1 cell up on the y-coordinate, 3 cells to the right on the x-coordinate and 2 cells down on the z-coordinate. Below, observe that those values are encoding using Zig-Zag encoding ($-1\rightarrow 1, 1 \rightarrow 2, -2 \rightarrow 3,\ldots$), and then they are packed in a 32-bit integer. 

Obviously, this works well as long as the assumption that there are small differences between two consecutive positions is maintained. However, there may be periods of time without information about the positions of the aircraft 
(for example, the aircraft is in an area without reception stations). In those cases, the 32-bit integer comprising $\langle x, y, z \rangle$ would not be enough. Observe that, to save space, our method does not explicitly store the time instant of a recorded position, it can be derived from its position inside the log. Therefore, 3DGraCT requires a method to manage that disappearances/appearances.

Between two consecutive snapshots ${\cal S}_{k}$ and ${\cal S}_{k+d}$, each object is represented by a log, ${\cal L}_{k,k+d}(id_j)$, where $id_j$ is the identifier of the object. It is a sequence of codewords of the following types: (1) an integer encoding a relative movement; (2) {\em Disappearance} ($D$) codeword, which means that we have no information about the position object $id_j$ from one time instant of ${\cal L}_{k,k+d}(id_j)$ until its end; (3) {\em Absolute appearance} ($AA$), which means that we have no information about the position of $id_j$ from the beginning of ${\cal L}_{k,k+d}(id_j)$ until a time instant covered by  ${\cal L}_{k,k+d}(id_j)$, where that information appears;    (4) {\em Relative disappearance}, which means that the information about the position of $id_j$ disappears in a time instant of  ${\cal L}_{k,k+d}(id_j)$, but reappears in a time instant of the same portion of the log.

\tolerance 10000 \pretolerance 10000

In order to maintain the synchronization of the sequences of values in ${\cal L}_{k,k+d}(id_j)$, the appearances and disappearances require the storage of their corresponding time instant. In addition, they also require the storage of the absolute position of the appearance/disappearance.  The relative disappearances imply the storage of the number of time instants they lasted and the relative movement with respect to the last known position.  

\tolerance 500 \pretolerance 500

In Figure \ref{log}(b), it is shown an example. The relative movements are depicted with the three relative displacements $\langle z,x,y \rangle$\footnote{$\langle z,x,y \rangle$ notation indicated that these three values are packed in a 32-bit integer.}. The array $\cal D$ stores the duration of a relative disappearance and the exact time instant of absolute appearances and disappearances. For example, in ${\cal L}_{0,4}(1)$, there is a relative disappearance that lasts two instants, and in ${\cal L}_{0,4}(7)$, the object appears at time instant 3. In addition, array $\cal P$ stores the relative movements of relative disappearances and the absolute position of absolute appearances or disappearances. For example,  in ${\cal L}_{0,4}(1)$, the $\langle 1,4,1\rangle$ tuple  in ${\cal P}_{0,4}(1)$ means that the object reappeared 1 cell upwards in the z-coordinate, 4 cells to the right in the x-coordinate, and 1 cell upwards in the y-coordinate.  In ${\cal L}_{0,4}(7)$, the object appears in the absolute position $(0,5,2)$ (see ${\cal P}_{0,4}(7)$). In the figure, the values are aligned to their corresponding time instants, but this is only for illustration purposes, thanks to the array $\cal D$, for one object, all the logs are stored as a sequence. $\cal D$  and $\cal P$ are compressed with DACs \cite{BLN13}, a compressor for sequences of integers that provides direct access to any position without the need of decompressing the previous numbers.

\medskip
\noindent
{\bf Compressing the log.}
Logs represent an important saving in space with respect to snapshots, but it is possible to obtain additional compression taking advantage the following fact: aircrafts spend most of the time following the same course at a constant speed. This situation will be represented in the logs as sequences of repetitive numbers, that is, the same relative displacements with respect to the previous time instant. 
These series of similar numbers are compressed very efficiently using a grammar compressor, such as Re-Pair.

To improve the query processing, the Re-Pair rules in 3DGraCT  are enriched with additional information. Each rule in $R$  has the following information: $s \rightarrow a, b, \#t, x, y, z, MBB$, where: (1) $s$, $a$ and $b$ are the components of a normal rule of Re-Pair, (2)  $\#t$ is the number of instants covered by the rule, (3) $\langle z,x,y \rangle$ are the relative coordinates of the final position of the object after the application of the rule (that is, the displacement considering (0,0,0) the initial position before the application of the rule)  and, (4) $MBB$ is the Minimum Bounding Box enclosing the movements of the rule. MBB is represented by six coordinates $(z1, x1, y1, z2, x2, y2)$, which are the  points at the ends of a diagonal of the box.

For example, in Figure \ref{log}, in ${\cal L}_{0,4}(4)$, the two $\langle 1,1,1\rangle$ consecutive relative movements   produce a rule, $W \rightarrow \langle 1,1,1\rangle ,\langle 1,1,1\rangle, 2, \langle 2,2,2\rangle, (0,0,0,2,2,2)$, and then  ${\cal L}_{0,4}(4)=W, D$.
Thanks to the additional information, the non-terminal symbols of the logs do not need to be decompressed in many cases. For example, if we wish to know the position of object 4 at $t_2$, we obtain its absolute position in the snapshot ${\cal S}_0$ (Figure \ref{snapshot}), which is (0,4,0), and then the first symbol of  ${\cal L}_{0,4}(4)$ ($W$) is applied. Since $W$ covers 2 time instants, its application to the position at $t_0$  produces the position of the object at the queried time instant. For this, the relative displacement $(2,2,2)$ is added to the original position, obtaining the position $(2,6,2)$.

\section{Querying}
\noindent
{\bf Obtain the position of an object.} 
To obtain the position of an object at a given time instant $t_q$, first, the algorithm retrieves the position of the object in the closest snapshot to $t_q$. If  the snapshot does not represent $t_q$, then the algorithm follows the movements through the log until it reaches $t_q$, as it was explained in the previous section, using the relative coordinates included in the rules when possible. When the nearest snapshot is located before $t_q$, the process follows a forward traversal of the log, otherwise, the process performs a backward traversal.

\medskip
\noindent
{\bf Obtain the trajectory of an object.} Given an interval of time $[t_s, t_e]$ and an object, this query obtains all the positions of the object between $t_s$ and $t_e$. First, the query obtains the position of the object at $t_s$  using the algorithm explained for the previous query, and then it applies the movements of the log until it reaches the position at $t_e$. Since the additional information of the rules does not contain the detailed positions of the trajectory, the algorithm has to decompress every non-terminal value of the log  containing a $ t_i \in [t_s, t_e]$.

\medskip
\noindent
{\bf Time slice query.}
Let $r = [x_1,y_1,z_1] \times [x_2,y_2,z_2]$ be a rectangular cuboid (or box) and $t_q$ a time instant, this query returns all objects within $r$ at $t_q$.  Let $(s_x, s_y, s_z)$ be the maximum speed vector of any object in our dataset, that is, the maximum speed in each of the three axes of the space achieved by any object in the dataset. We denote $E_r(t_k,t_q)$, the \textit{expanded region} of $r$ from $t_k$ to $t_q$, as the area that contains any  object active at $t_k$  capable of being located within $r$ at $t_q$. Hence, $E_r(t_k,t_q)$ is $r$ extended in the three dimensions; in the x-axis to the coordinates $[x_1-s_x\cdot(t_q-t_k),x_2+s_x\cdot(t_q-t_k)]$, and repeat the same for the y-axis and z-axis. Assuming that the closest snapshot is ${\cal S}_{k}$ and that $t_k \leq t_q$, the algorithm obtains the candidate objects ${\cal C}$ inside $E_r(t_k,t_q)$ at $t_k$. If $t_k=t_q$, the algorithm returns ${\cal C}$. Otherwise, it tracks the movements in ${\cal L}_{k, k+d}$ for each object in ${\cal C}$ until it reaches $t_q$. During this process, after obtaining the position of an object $c_j$ at $t_i$, we can discard $c_j$ if it is outside $E_r(t_i, t_q)$. The position at $t_q$ can be given by a terminal or a non-terminal value. In the first case, we apply the movement and check if the object is within $r$. In the second case, the object is part of the solution when the MBB of the additional information of the rule defining the non-terminal value is completely contained in $r$, and the object can be pruned if its MBB does not intersect $r$. However, when the MBB intersects  $r$ (but it is not completely contained), the algorithm has to decompress the non-terminal symbol using the Re-Pair rule to obtain the exact position of the object at $t_q$. If the closest snapshot to $t_q$ is after it, then the algorithm  performs the same process backwards. 

\medskip
\noindent
{\bf Time interval query.}
Given a box $r$ and an interval of time $[t_s, t_e]$, this query obtains all objects within $r$ at any $t_i \in [t_s, t_e]$. This query could be solved in a similar way to the previous one. However, to avoid large expanded regions, that lead to track too many  candidate objects,
  the query interval $[t_s, t_e]$ is divided into as many queries as portions of log overlaps. Then, each one of these portions 
$[t^\prime_{s}, t^\prime_{e}]$ can be solved in a similar way to time-slice. First, the algorithm obtains the candidates from the closest snapshot, using the expanded region with respect to $t^\prime_{e}$; then it applies the movements of the log. During the processing of the log  of a candidate object $c_j$, the algorithm has to take into account that when the traversal reaches a symbol $s_m$ that after its application obtains the position $(x_i, y_i, z_i)$ at a time instant $ t_i \in [t_s, t_e]$: 
(1)  $c_j$ is part of the solution if $(x_i, y_i, z_i)$ is within $r$;
(2)  if $(x_i, y_i, z_i)$ is not within $E_r(t_i,  t^\prime_{e})$, then $c_j$ can be discarded of the processing of the current portion;
(3) if $(x_i, y_i, z_i)$   is outside $r$ but within $E_r(t_i,  t^\prime_{e})$, then $c_j$ continues as a candidate that needs to be tracked.
If $s_m$ is a non-terminal symbol that produces a position at $t_i>t^\prime_{e}$ and covers the time interval $[t_u, t_i]$, where $t_u \leq t^\prime_{e}$:
(1)  if the MBB of $s_m$  is fully within  $r$, then $c_j$ is part of the solution
(2)  if the MBB  of $s_m$ does not intersect  $r$, then  $c_j$ is discarded in the processing of the current portion.
(3)  if the MBB of $s_m$  intersects  $r$, the algorithm has to decompress $s_m$ to check if $s_m$  involves any $t_l \in [t_u, t^\prime_{e}]$ whose position is within $r$.


\section{Experimental Evaluation}

Our experiments analyze space/time tradeoffs of 3DGraCT using real-world ADS-B data. We also evaluate the use of interpolation to fill in large periods of missing data during the trajectory. For comparison purposes, we propose a baseline including the MVR-tree \cite{PapadiasT01}, but we do not include ADS-BI \cite{8283602} because it does not provide altitude-based queries, and its inner index stores some string dimensions 
which are not covered by 3DGraCT. 

Both 3DGraCT and the MVR-tree are coded in C++. 3DGraCT uses some structures from SDSL \cite{gbmp2014sea} and MVR-tree is obtained from the spatialindex library (\texttt{libspatialindex.github.io}). All experiments were run on an Intel\textsuperscript{\textregistered} Core\textsuperscript{TM} i7-3820 CPU@3.60GHz (4 cores), 10MB of cache and 64GB of RAM, over Ubuntu 12.04.5 LTS (kernel 3.2.0-115, 64 bits), using gcc 4.6.4 with \texttt{-O9} flag.

\medskip
\noindent
{\bf Dataset details.}
We use four real ADS-B datasets including descriptive data of flights between different airports of Europe (see details in Appendix \ref{details}). Each dataset covers a different period of time, namely {\em one day}, {\em one week}, {\em two weeks}, and {\em one month}. Positions are discretized into a cube where the cell size is 5 kilometers in x-axis, 5 kilometers in y-axis, and 100 meters in z-axis.  Since aircraft positions can contain incorrect information and they can be emitted at different time rates, we discard incorrect positions and normalize timestamps to obtain regular instants every 15 seconds.

Gate-to-gate trajectories are difficult to reconstruct from ADS-B data because some broadcasted positions are lost, mainly due to lack of coverage. Although we use disappearance and reappearance codewords to represent these situations, we consider relevant to understand how they affect to 3DGraCT tradeoffs. We use the original datasets to generate a new ones, where aircraft positions are interpolated when no information is available during, at least, 15 minutes.
As consequence, we have eight datasets: four real-world datasets (\textit{1D}, \textit{1W}, \textit{2W}, \textit{1M}) and four interpolated datasets (\textit{1D-I}, \textit{1W-I}, \textit{2W-I}, \textit{1M-I}). Table \ref{table:dataset} shows the details of each dataset. Note that the fourth and fifth rows  give, respectively, dataset sizes of binary and p7zip-compressed representations.

\begin{table}[t]
 \scriptsize
 \centering
 \begin{tabular}{|l|p{0.1\textwidth}|p{0.1\textwidth}|p{0.1\textwidth}|p{0.1\textwidth}|p{0.1\textwidth}|p{0.1\textwidth}|p{0.1\textwidth}|p{0.1\textwidth}|}
  \hline
  \textbf{Dataset} & \textbf{1D} & \textbf{1D-I} & \textbf{1W} & \textbf{1W-I} & \textbf{2W} & \textbf{2W-I} & \textbf{1M} & \textbf{1M-I}\\
  \hline 
  Time & 1 day & 1 day & 1 week & 1 week & 2 weeks & 2 weeks & 1 month & 1 month \\
  Objects & 1082& 1082& 1764& 1764& 2003& 2003& 2263& 2263\\
  Interpolated & No & Yes & No & Yes & No & Yes & No & Yes \\
  \hline
    Binary & 7.31M & 7.68M & 55.32M & 58.27M & 115.57M & 122.03M & 261.01M& 275.35M\\
    \hline
  p7zip & 1.71M & 1.86M & 12.58M & 13.09M & 26.03M & 27.18M & 57.45M & 60.14M\\
  (ratio)  & 23.41\% & 24.19\% & 22.73\% & 22.47\% & 22.53\% & 22.27\% & 22.01\% & 21.84\% \\
  \hline
 \end{tabular}
 \caption{Dataset details.}
 \label{table:dataset}
\end{table}

\begin{figure}[p]
	\centering     
\subfigure[\texttt{Compression ratio}]
{\label{fig:comp}\includegraphics[width=0.45\textwidth]{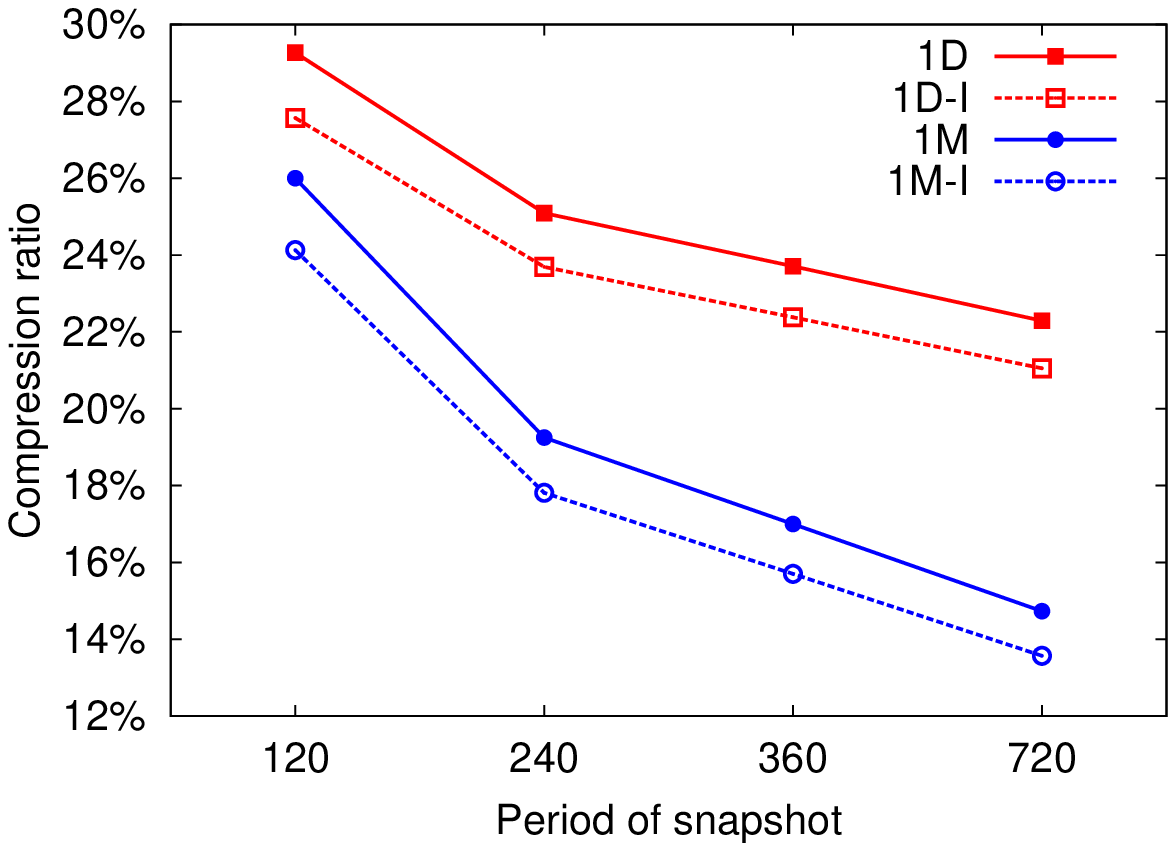}}
\subfigure[\texttt{Object $t$}]
{\label{fig:time-ot}\includegraphics[width=0.45\textwidth]{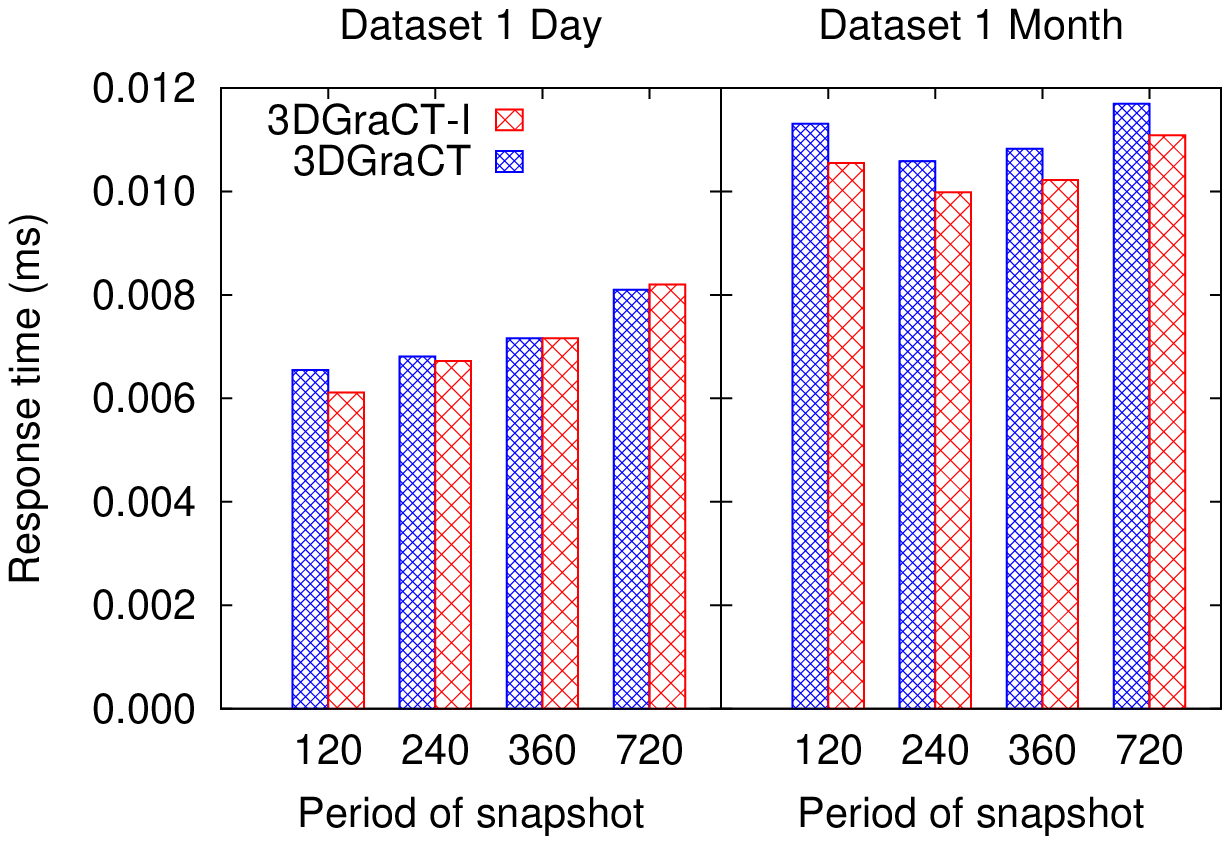}}
\subfigure[\texttt{Trajectory}]
{\label{fig:time-trajectory}\includegraphics[width=0.45\textwidth]{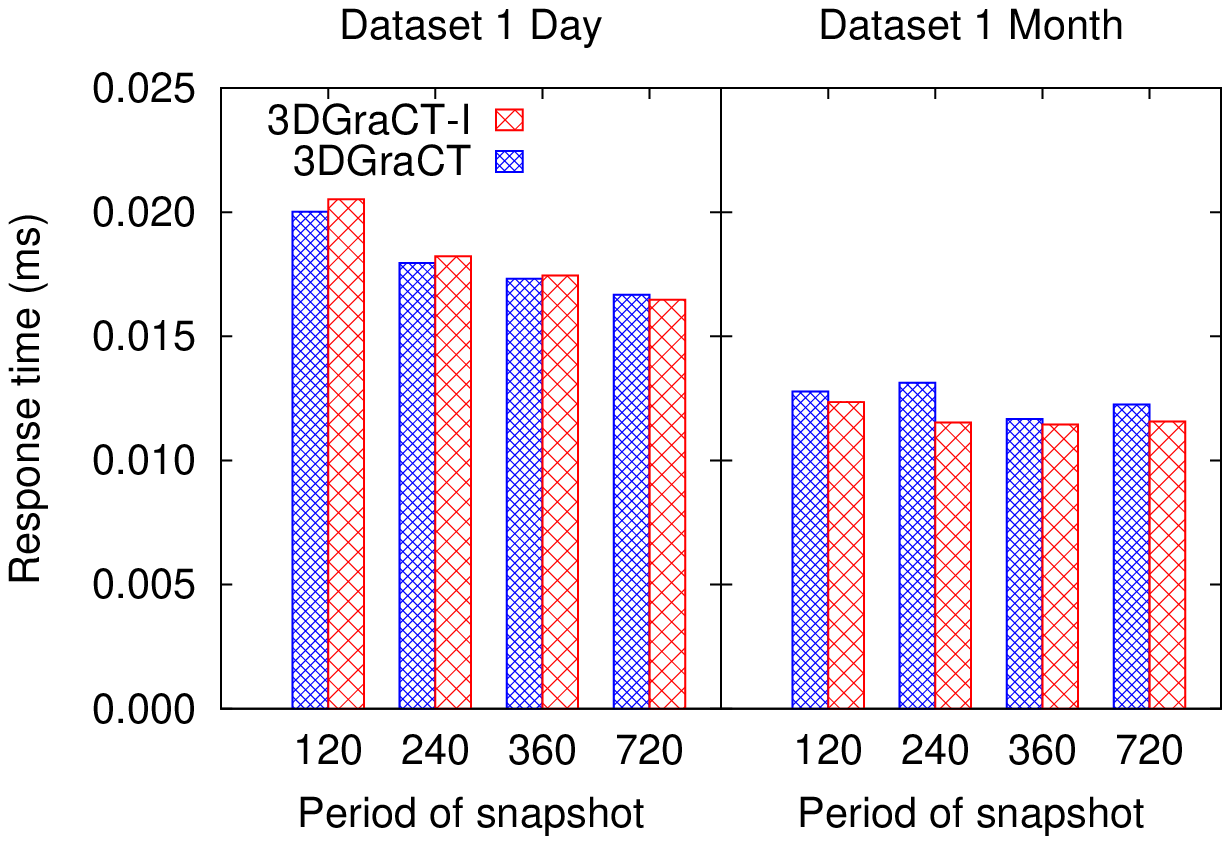}}
\subfigure[\texttt{Time Slice Small}]
{\label{fig:time-sliceS}\includegraphics[width=0.45\textwidth]{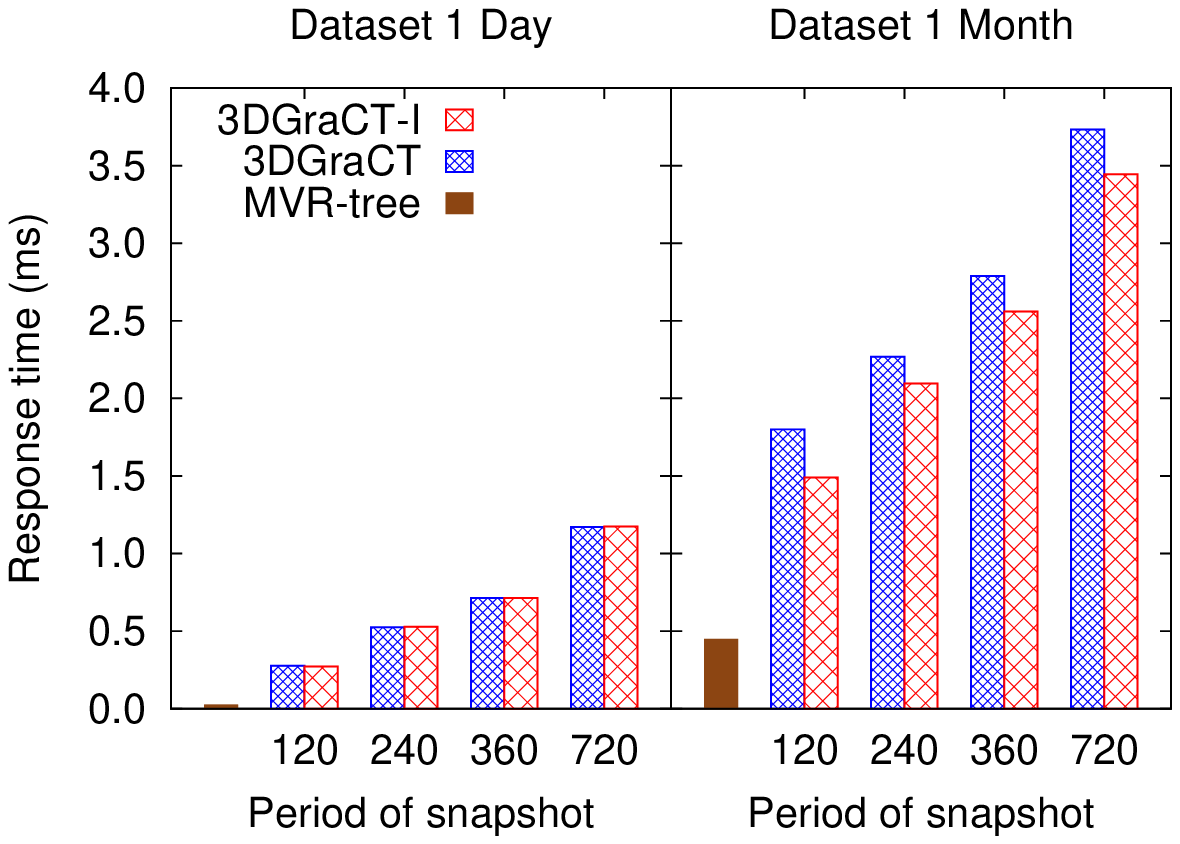}}
\subfigure[\texttt{Time Slice Large}]
{\label{fig:time-sliceL}\includegraphics[width=0.45\textwidth]{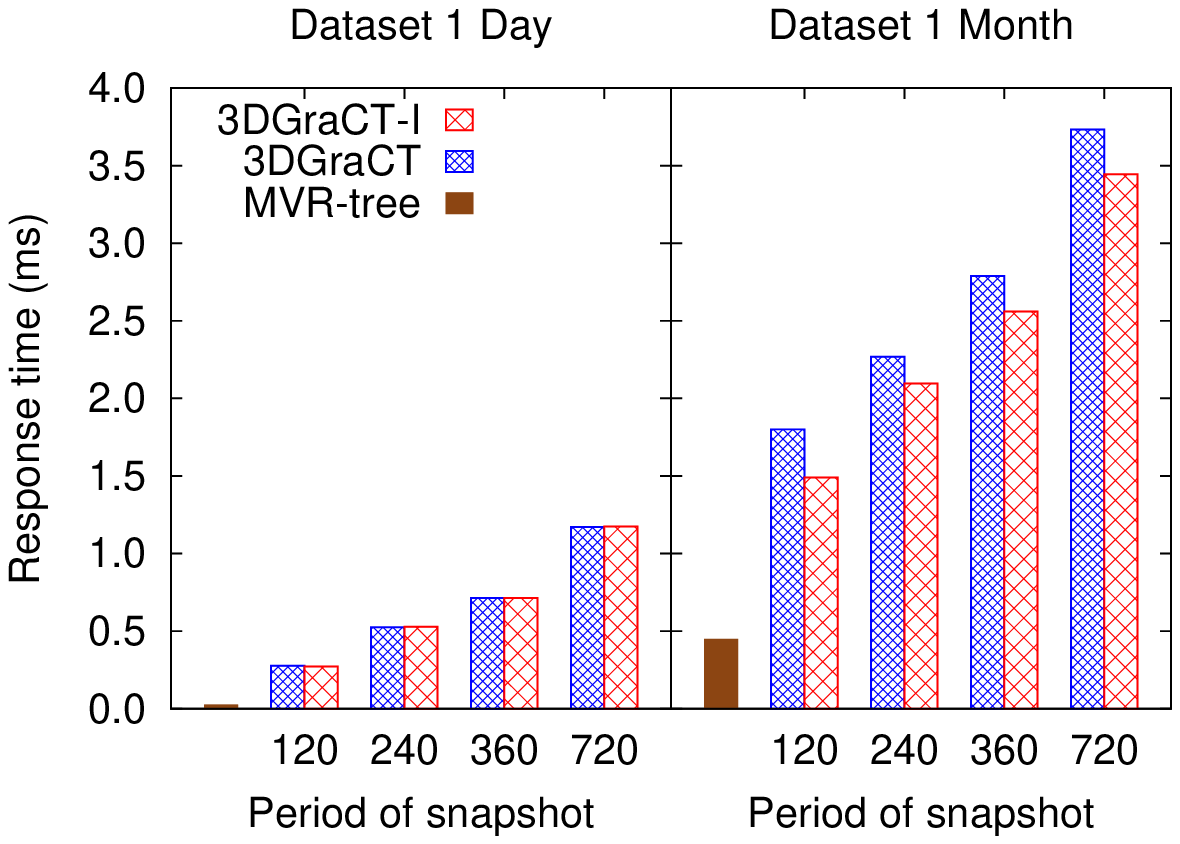}}
\subfigure[\texttt{Time Interval Small}]
{\label{fig:time-intervalS}\includegraphics[width=0.45\textwidth]{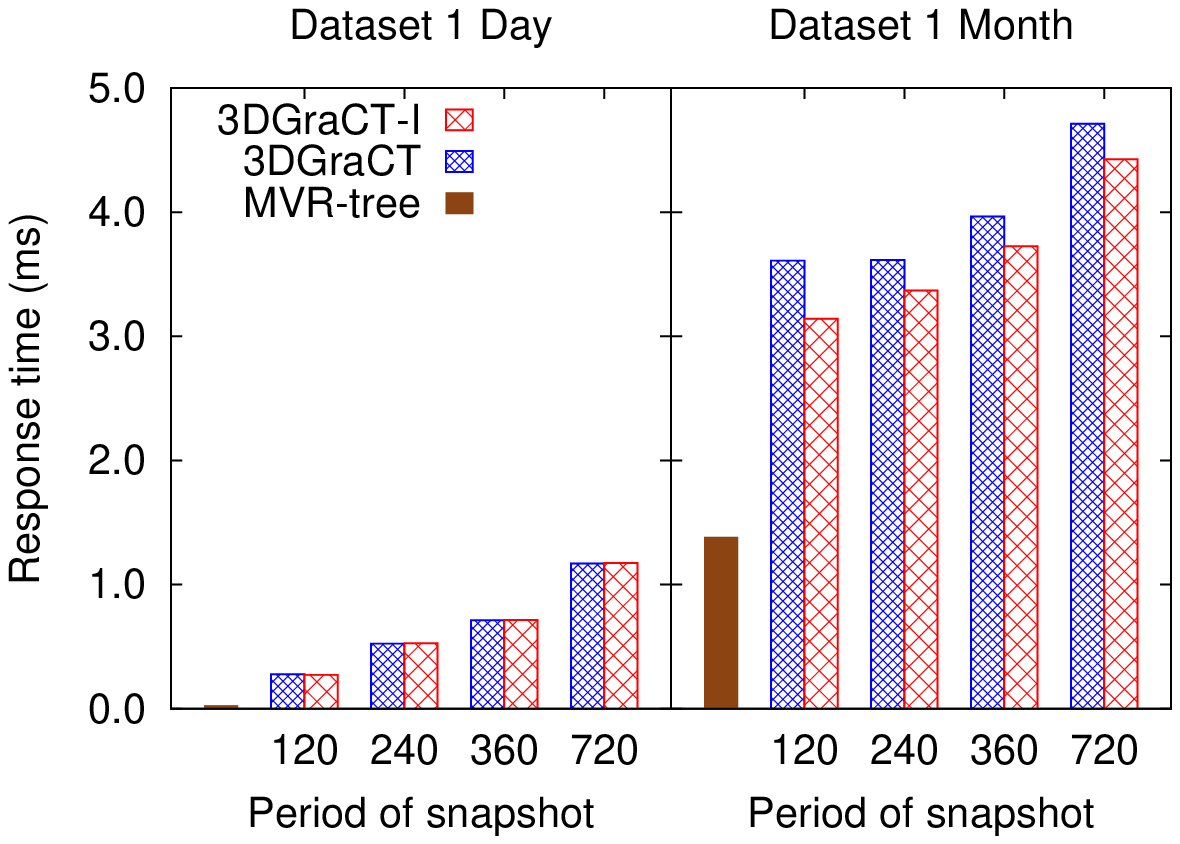}}
\subfigure[\texttt{Time Interval Large}]
{\label{fig:time-intervalL}\includegraphics[width=0.45\textwidth]{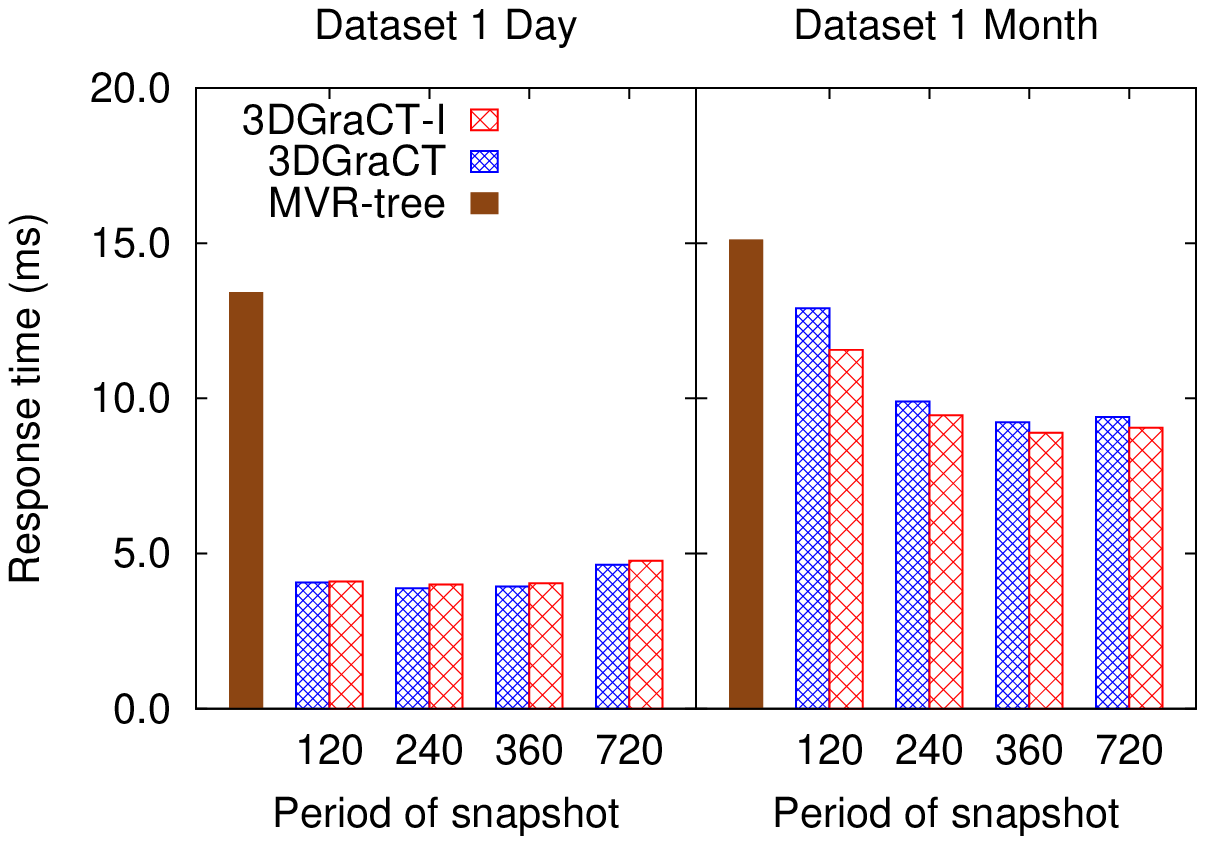}}
\subfigure[\texttt{Growing time interval}]
{\label{fig:interval}\includegraphics[width=0.45\textwidth]{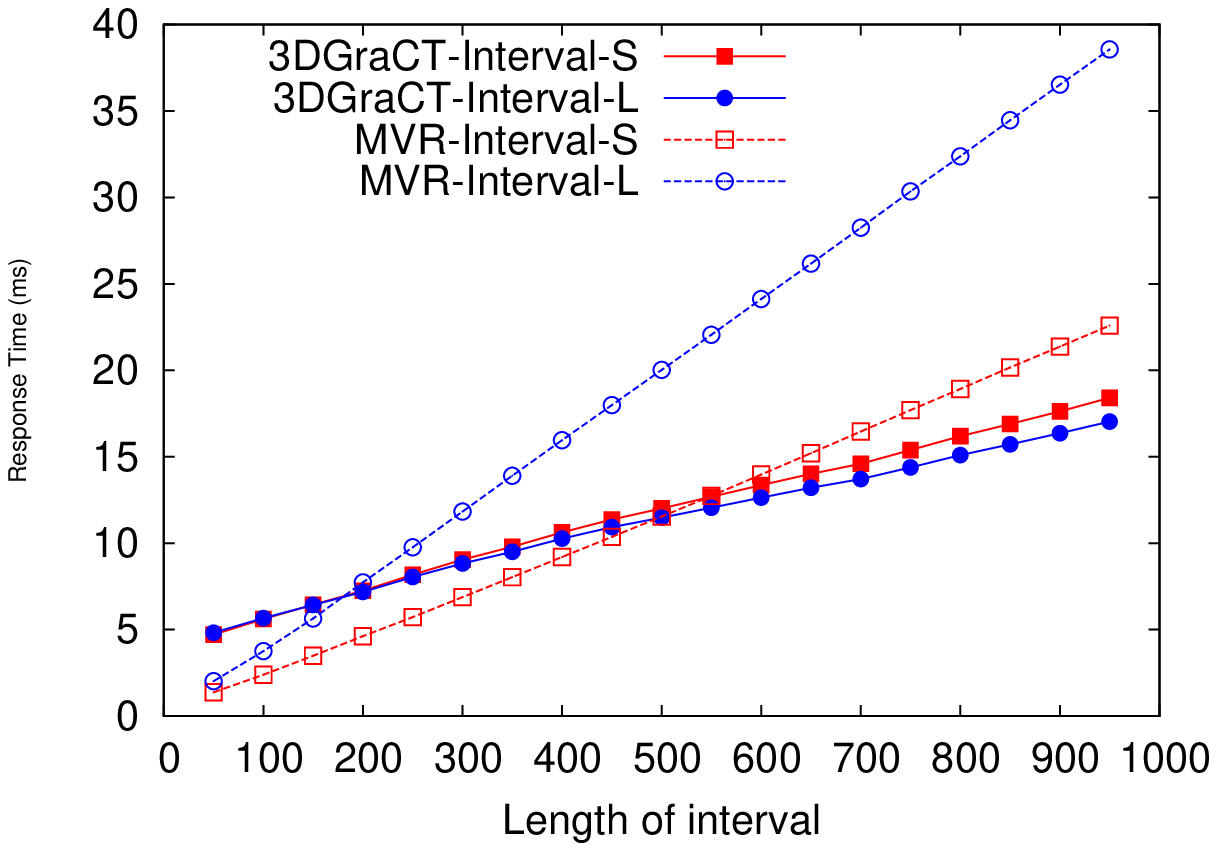}}
\caption{Compression ratio and query times (ms).}
\label{fig:experiment2}
\end{figure}

\medskip
\noindent
{\bf Compression ratio.}  We define compression ratio as the ratio between the binary size and the compressed size. The last row of Table \ref{table:dataset} gives compression ratios reported by p7zip for all datasets, while Figure \ref{fig:comp} illustrates 3DGraCT numbers for {\em one day} and {\em one month} datasets, using different periods of snapshot (120, 240, 360 and 720 time instants). p7zip report stable ratios around 22-24\%, but 3DGraCT effectiveness is clearly influenced by the distance between snapshots, because snapshot encoding requires more space than log compression. Thus, the more-distanced the snapshots are, the better the results are. In our experiments, 3DGraCT reports its best ratios using a separation of 720 time instants between snapshots, outperforming p7zip in all datasets. For instance, 3DGraCT reports 22.29\% for \textit{1D} and p7zip 23.41\%. This gap increases for larger datasets: 3DGraCT only needs 14.73\% of the original \textit{1M} size, while p7zip demands 22.01\%. Thus, 3DGraCT is more effective than a powerful compressor like p7zip, while retaining search capabilities.

This comparison also applies for interpolated datasets. Note that, in this case, 3DGraCT reports slightly better results, meaning that missing information adds an small overhead ($\approx$ 2\%) to our structure.


\medskip
\noindent
{\bf Query times.}
Query times are averaged over the following settings:
(1) \textit{Object t}: {\em 20{,}000 queries} that obtain the position of an object at a given time instant, (2) \textit{Trajectory}: {\em 10{,}000 queries} obtaining trajectories that cover 2{,}000 time instants, (3) \textit{Time Slice S}:  {\em 1{,}000 time slice queries} involving a small region ($20 \times 20 \times 20$), (4) \textit{Time Slice L}:  {\em 1{,}000 time slice queries} specifying large regions ($160 \times 160 \times 160)$, (5) \textit{Time Interval S}: {\em 1{,}000 time interval queries} involving small regions and intervals of 50 time instants, (6) \textit{Time Interval L}: {\em 1{,}000 time interval queries} specifying large regions and intervals of 400 time instants. Query times for 3D-GraCT over real-world (3D-GraCT) and interpolated (3D-GraCT-I) datasets are distinguished in the following figures.

Figure \ref{fig:time-ot} shows that query times of \textit{Object t} increase with distance between snapshots  because larger log portions must be processed.
 On the contrary, Figure \ref{fig:time-trajectory} shows that \textit{Trajectory} queries are slowler for less distanced snapshots because more snapshots must be checked.

In region queries, Time Slice and Time Interval, the number of candidates
depends on the period between snapshots.  \textit{Time Slice} is slower as the distance between snapshot gets larger (see Figures \ref{fig:time-sliceS} and \ref{fig:time-sliceL}), because the extended region grows and the number of candidates that are tracked is also larger. Figures \ref{fig:time-intervalS} and \ref{fig:time-intervalL} show that \textit{Time Interval} queries  behave similar to 
Time Slice ones, except in the right part of Figure \ref{fig:time-intervalL}. In this case, the expanded region covers the whole space for each period of snapshot, so the number of candidates between different settings remains constant. Thus, traversing the log demands the same computation, but less snapshots are checked for larger periods. 

 Finally, it is worth noting that the effect of interpolation is not very relevant to 3DGraCT performance. It is only a slight improvement for region queries and large datasets. Thus, we conclude that the interpolation of missing positions avoids the cost of managing appearances and reappearances, improves Re-Pair effectiveness, and allows logs to be processed faster. For this reason, querying real-world datasets are 3\%-10\% slower.

\medskip
\noindent
{\bf Comparison with MVR-tree.}
3DGraCT and MVR-tree are compared over the real-world datasets of our setup: \textit{1D}, \textit{1W}, \textit{2W} and \textit{1M}. It is worth noting that MVR-Tree space requirements are 250-300 times larger than 3DGraCT one, but we tune MVR-Tree to run on main-memory.


Our analysis show that MVR-tree is only efficient for Time Slice, Time Interval, and knn queries.
Although MVR-tree can obtain the position of an object at a given time instant, or can follow the trajectory of the object in a given interval, these are expensive queries. 

MVR-tree can be enhanced with an auxiliary 3DR-tree \cite{PapadiasT01}, but the resulting structure would consume even more space. Thus, we only analyze queries where MVR-tree is efficient.


Figures \ref{fig:time-sliceS} and \ref{fig:time-sliceL} show that MVR-tree outperforms 3DGraCT in Time Slice queries. However, our structure is better in Time Interval queries for large intervals (Figure \ref{fig:time-intervalL}).
We study the turning point where the 3DGraCT starts to improve the MVR-tree, by increasing the time interval length. Figure \ref{fig:interval} shows this comparison for the {\em 1M} dataset, and a period of snapshot of 720. 3DGraCT outperforms MVR-tree for time intervals over 550 and 200 time instants in small and large regions, respectively.

\section{Conclusions}

This paper introduces 3DGraCT, a new data structure capable of compressing and querying 3D trajectories with no prior decompression. 3DGraCT extends an existing 2D compact data structure (GraCT) to support a third dimension, enabling object positions to be enhanced with descriptive altitude data. Our improvements to GraCT are more than just improving object descriptions because 3DGraCT also enables for resolving altitude-based queries. 

3DGraCT has been evaluated using real-world trajectories reconstructed from ADS-B descriptions. 3DGraCT reports better compression ratios than universal compressors like p7zip (3DGraCT uses up to 50\% less space), while retaining search capabilities. Compared to traditional spatio-temporal solutions, 3DGraCT needs 2 orders of magnitude less space than MVR-tree, being competitive in query performance. Finally, we also study the effect of missing subtrajectories, concluding that interpolation is effective in different cases.

\bibliographystyle{splncs03}
\bibliography{bibliografia}

\appendix
\section{Appendix}\label{details}
The datasets used in our experimenation have been obtained from the OpenSky Network\footnote{\texttt{https://opensky-network.org/}}. We have chosen ADS-B messages broadcasted by aircrafts of 30 different airlines and describe flights between 30 European airports:

\begin{itemize}
    \item Airlines (ICAO code): {\tt AEA, AEE, AFR, AUA, AZA, BAW, BEE, BEL, BER, DLH, EIN, EWG, EZS, EZY, FDX, FIN, GWI, IBE, IBK, IBS, KLM, LOT, NAX, NLY, RYR, SAS, SHT, SWR, TAP}, and {\tt VLG}.
    \item Airports (ICAO code): {\tt EBBR, EDDF, EDDK, EDDL, EDDM, EDDT, EFHK, EGCC, EGKK, EGLL, EGPH, EGSS, EHAM, EIDW, EKCH, ENGM, EPWA, ESSA, LEBL, LEMD, LEPA, LFPG, LFPO, LGAV, LIMC, LIRF, LOWW, LPPT, LSGG}, and {\tt LSZH}. 
\end{itemize}

ADS-B messages were captured from 2017-01-02 to 2017-01-31, and sampled as follows:

\begin{itemize}
  \item 1day :  {\tt 2017-01-02}.
  \item 1week:  {\tt 2017-01-02 -- 2017-01-08}.
  \item 2weeks: {\tt 2017-01-02 -- 2017-01-15}.
  \item 1month: {\tt 2017-01-02 -- 2017-01-31}.
\end{itemize}

\end{document}